\documentclass[12pt]{article}

\usepackage{psfig}
\usepackage{amssymb}
\usepackage{amsmath}

\def\lsim{\
  \lower-1.2pt\vbox{\hbox{\rlap{$<$}\lower5pt\vbox{\hbox{$\sim$}}}}\ }
\def\gsim{\
  \lower-1.2pt\vbox{\hbox{\rlap{$>$}\lower5pt\vbox{\hbox{$\sim$}}}}\ }

\makeatletter

\@addtoreset{equation}{section}

\makeatother 

\begin{document}

\date{\empty}

\title{Braneworld cosmological solutions and their stability}
\author{Dmytro Iakubovskyi$^{1,2}$\footnote{E-mail:
yakubovsky@bitp.kiev.ua} \ and Yuri Shtanov$^{2}$\footnote{E-mail:
shtanov@bitp.kiev.ua}
\medskip \\
{\small \it $^1$Department of Physics, Taras Shevchenko National University,
          Kiev 03022, Ukraine } \\
{\small \it $^2$Bogolyubov Institute for Theoretical Physics, Kiev 03143,
Ukraine}}

\maketitle

\begin{abstract}
We consider cosmological solutions and their stability with respect to
homogeneous and isotropic perturbations in the braneworld model with the
scalar-curvature term in the action for the brane. Part of the results are
similar to those obtained by Campos and Sopuerta for the Randall--Sundrum
braneworld model.  Specifically, the expanding de~Sitter solution is an
attractor, while the expanding Friedmann solution is a repeller.  In the
braneworld theory with the scalar-curvature term in the action for the brane,
static solutions with matter satisfying the strong energy condition exist not
only with closed spatial geometry but also with open and flat ones even in the
case where the dark-radiation contribution is absent. In a certain range of
parameters, static solutions are stable with respect to homogeneous and
isotropic perturbations.
\end{abstract}

\noindent PACS number(s): 04.50.+h

\section{ Introduction }

Braneworld models have become quite popular in the high-energy and
gravita\-tional phy\-sics during last several years. There are at least three
reasons for this. Firstly, most of the brane models, in particular, the
Randall--Sundrum model \cite{RS}, are inspired directly by the popular string
theory (specifically, by the Ho\u{r}ava--Witten model \cite{HW}) and represent
an interesting alternative for compactification of extra spatial dimensions.
Secondly, there remains a possibility of solving the mass-hierarchy problem in
the context of large extra dimensions (see \cite{ADD,RS}). And thirdly, it has
been shown that models with non-compact extra dimensions and branes can be
consistent with the current gravity experiments.

Braneworld models also turned out to be consistent with modern cosmology, at
the same time exhibiting new specific features. First of all, this concerns the
braneworld models of ``dark energy,'' or of the currently observed acceleration
of the universe (see \cite{SS} in this respect), but this is also true with
respect to the early stages of cosmological evolution such as inflation (see,
e.g., \cite{Maartens}).

A broad class of cosmological solutions in braneworld theory was systematically
analyzed by Campos and Sopuerta \cite{CS} using convenient phase space
variables similar to those introduced in \cite{GE}.  These authors gave a
complete description of stationary points in an appropriately chosen phase
space of the cosmological setup and investigated their stability with respect
to homogeneous and isotropic perturbations.  The authors worked in the frames
of the Randall--Sundrum braneworld theory without the scalar-curvature term in
the action for the brane.  In this paper, we are going to extend this analysis
by investigating the case where this term is present in the action.

The scalar-curvature term for the brane was introduced in \cite{DGP} (see also
\cite{CHS}) as a method of making gravity on the brane effectively
four-dimensional even in the flat infinite bulk space. The corresponding
cosmological models were initiated in \cite{CHS,DDG}.  The necessity of this
term on the brane arises when one considers the complete effective action of
the theory (see, e.g., \cite{HHR}).  From this viewpoint, the term with brane
tension represents the zero-order term in the expansion of the total action for
the brane in powers of curvature of the induced metric, while the next term in
the action for the brane is exactly the scalar-curvature Hilbert--Einstein
term.  From another (complementary) viewpoint \cite{DGP}, this contribution is
induced as a quantum correction to the effective action for the brane gravity
after one takes into account the quantum character of the matter confined to
the brane.

In this paper, we will show that the presence of the scalar-curvature term in
the action for the brane enlarges the class of cosmological solutions, in
particular, it allows for the existence of static cosmological solutions with
ordinary matter content even in the case of spatially flat or open cosmology, a
specific feature of the braneworld cosmology that was also noted in
\cite{CS,GM,SS}.  In the Randall--Sundrum model, flat or open static universe
with ordinary matter requires the presence of the negative dark-radiation term,
while, in the model with scalar-curvature term in the action for the brane to
be discussed in this paper, the presence of dark radiation may be not
necessary.

The paper is organized as follows: \  In Sec.~\ref{basic}, we introduce the
basic equations describing the theory with one brane in the five-dimensional
bulk space. Then, in Sec.~\ref{static}, we introduce a method for the
investigation of static solutions. Using this method, we describe both the
general-relativity case and the case of braneworld theory that contains the
scalar-curvature term in the action for the brane. In the partial case of
spatially flat universe, we determine the range of parameters where static
solutions exist and where they are stable with respect to homogeneous and
isotropic perturbations. In Sec.~\ref{nonstatic}, we develop an approach which
is used to describe non-static solutions, and apply it to the braneworld theory
under investigation. In Sec.~\ref{static-case}, we return to the issue of
static solutions with arbitrary spatial curvature using conveniently chosen
variables. Finally, in Sec.~\ref{conclusions}, we present our conclusions.

\section{Basic equations}\label{basic}

In this paper, we consider the case where the braneworld is the time-like
boundary of a five-dimensional purely gravitational Lorentzian space (bulk),
which is equivalent to the case of a brane embedded in the bulk with
$\mathbb{Z}_2$ symmetry of reflection with respect to the brane. The theory is
described by the action \cite{CHS}:
\begin{equation}\label{act}
S=-M^3\left[\int_{\rm bulk}(\mathcal{R}-2\Lambda_{\rm b})-2\int_{\rm brane}
K\right]-\int_{\rm brane}(m^2R-2\sigma)+\int_{\rm brane} L(h_{ab},\phi)\, .
\end{equation}
Here, $\mathcal{R}$ is the scalar curvature of the metric $g_{ab}$ in the
five-dimensional bulk, and $R$ is the scalar curvature of the induced metric
$h_{ab}=g_{ab}-n_{a}n_{b}$ on the brane, where $n^{a}$ is the vector field of
the inner unit normal to the brane, and the notation and conventions of
\cite{Wald}  are used. The quantity $K=K_{ab}h^{ab}$ is the trace of the
symmetric tensor of extrinsic curvature $K_{ab}=h^{c}_{a}\nabla_{c}n_{b}$ on
the brane. The symbol $L(h_{ab}, \phi)$ denotes the Lagrangian density of the
four-dimensional matter fields $\phi$ whose dynamics is restricted to the brane
so that they interact only with the induced metric $h_{ab}$. All integrations
over the bulk and brane are taken with the natural volume elements
$\sqrt{-g}d^{5}x$ and $\sqrt{-h}d^4x$, respectively, where $g$ and $h$ are the
determinants of the matrices of components of the corresponding metrics in a
coordinate basis. The symbols $M$ and $m$ denote, respectively, the five- and
four-dimensional Planck masses, $\Lambda_{\rm b}$ is the bulk cosmological
constant, and $\sigma$ is the brane tension.

The term containing the scalar curvature of the induced metric on the brane
with the coupling $m^2$ in action (\ref{act}) was originally absent from
braneworld models. However, very soon it became clear that this term is
qualitatively essential for describing the braneworld dynamics and, moreover,
it is inevitably generated as a quantum correction to the matter action in
(\ref{act}) --- in the spirit of an idea that goes back to Sakharov
\cite{Sakharov} (see also \cite{BD}). Note that the effective action for the
brane typically involves an infinite number of terms of higher order in
curvature (this was pointed out in \cite{CHS} for the case of braneworld
theory; a similar situation in the context of the AdS/CFT correspondence is
described in \cite{HHR}).  In this paper, we retain only the terms linear in
curvature. The effects of the curvature term on the brane in linear
approximation were studied in \cite{DGP,CH}, where it was shown that it leads
to four-dimensional law of gravity on sufficiently small scales. The presence
of this term also leads to qualitatively new features in cosmology, as
demonstrated, in particular, in \cite{CHS,SS}. For some recent reviews of the
results connected with the induced-gravity term in the brane action, one may
look into \cite{review}.

Variation of action (\ref{act}) gives us the equations describing the dynamics
in the bulk
\begin{equation}\label{eq_mot1}
\mathcal{G}_{ab}+\Lambda_{\rm b}g_{ab}=0\, ,
\end{equation}
and on the brane
\begin{equation}\label{eq_mot2}
m^2G_{ab}+\sigma h_{ab}=\tau_{ab}+M^3(K_{ab}-Kh_{ab})\, .
\end{equation}
The second equation generalizes the Israel junction conditions \cite{Israel}
and the corresponding equation of the Randall--Sundrum model to the presence of
the scalar-curvature term on the brane ($m \ne 0$).

One can consider the Gauss identity
\begin{equation}\label{Gauss}
R_{abc}{}^d = h_a{}^f h_b{}^g h_c{}^k h^d{}_j {\cal R}_{fgk}{}^j + K_{ac}
K_b{}^d - K_{bc} K_a{}^d \, ,
\end{equation}
and, following the original procedure of \cite{SMS}, contract it once on the
brane using equations (\ref{eq_mot1}) and (\ref{eq_mot2}). In this way, one
obtains the effective equation on the brane that generalizes the result of
\cite{SMS} to the presence of the brane curvature term:
\begin{equation}\label{effective}
G_{ab} + \Lambda_{\rm eff} h_{ab} = \frac{1}{M_{\rm P}^2 } \tau_{ab} +
\frac{1}{1 + 2 \sigma m^2 /3 M^6 } \left( M^6 Q_{ab} - W_{ab} \right) \, ,
\end{equation}
where
\begin{equation}
\Lambda_{\rm eff} = \frac{\Lambda_{\rm RS}}{1 + 2 \sigma m^2 / 3 M^6} \, ,
\quad \Lambda_{\rm RS} = \frac{\Lambda_{\rm b}}{2} + \frac{\sigma^2}{3 M^6} \,
, \quad M_{\rm P}^2 = m^2 + \frac{3 M^6}{2 \sigma} \, ,
\end{equation}
\begin{equation}
Q_{ab} = \frac13 E E_{ab} - E_{ac} E^{c}{}_b + \frac12 \left(E_{cd} E^{cd} -
\frac13 E^2 \right) h_{ab}
\end{equation}
is the quadratic expression with respect to the `bare' Einstein equation
$E_{ab} \equiv m^2 G_{ab} - \tau_{ab}$ on the brane, $E = h^{ab} E_{ab}$, and
$W_{ab} \equiv h^c{}_a h^e{}_b W_{cdef} n^d n^f$ is a projection of the bulk
Weyl tensor $W_{abcd}$ to the brane. The limit of $m \to 0$ leads us to the
original result of \cite{SMS}.

Contracting Eq.~(\ref{effective}) once again, one obtains the following scalar
equation on the brane:
\begin{equation}\label{2nd}
M^{6}(R-2\Lambda_{\rm b})+(m^2G_{ab}+\sigma h_{ab}-\tau_{ab})(m^2G^{ab}+\sigma
h^{ab}-\tau^{ab})-\frac{1}3(m^2R-4\sigma+\tau)^2=0\, ,
\end{equation}
where $\tau=\tau_{ab}h^{ab}$. It is argued in \cite{Shtanov} that, if one is
interested only in the evolution on the brane and if no additional boundary or
regularity conditions are specified in the bulk (as often is the case), then
one is left only with this scalar equation.  Having obtained a solution of
(\ref{2nd}), one can integrate the gravitational equations in the bulk using,
for example, Gaussian normal coordinates \cite{SMS,Shtanov}.

The brane filled by a homogeneous and isotropic ideal fluid is described by the
Fried\-mann--Robertson--Walker (FRW) metric
\begin{equation}
ds^2=-dt^2+a^2(t)\left[\frac{dr^2}{1-\kappa r^2}+r^2d\Omega^2\right]\, ,
\end{equation}
where $\kappa=-1$, $0$, or $+1$. The stress--energy tensor of the ideal fluid
takes the form
\begin{equation}
\tau_{ab}=(\rho+p)u_{a}u_{b}+ph_{ab} \, ,
\end{equation}
where $\rho$ is the energy density, and $p$ is the pressure.  In this paper, we
assume a simple equation of state $p = w \rho$, where $w$ is constant.

Equation (\ref{2nd}) leads to the following cosmological equation in the
braneworld theory:
\begin{equation}\label{second}
9M^{6}\left[\frac{\ddot{a}}{a}+H^2+\frac{\kappa}{a^2}-\frac{\Lambda_{\rm
b}}3\right] +\left[\rho(1+3w)-2\sigma+6m^2\frac{\ddot{a}}{a}\right]
\left[\rho+\sigma - 3m^2\left(H^2+\frac{\kappa}{a^2}\right)\right]=0\, ,
\end{equation}
where $H \equiv \dot a / a$ is the Hubble parameter. After multiplying this
equation by $a^3\dot{a}$ and using the conservation of energy
\begin{equation}
\dot{\rho}+3H(\rho+p)=0 \, ,
\end{equation}
one can integrate Eq.~(\ref{second}) and obtain the first-order differential
equation \cite{Shtanov,SS}
\begin{equation}\label{1st}
m^4\left(H^2+\frac{\kappa}{a^2}-\frac{\rho+\sigma}{3m^2}\right)^2=
M^{6}\left(H^2+\frac{\kappa}{a^2}-\frac{\Lambda_{\rm
b}}{6}-\frac{C}{a^4}\right)\, ,
\end{equation}
where $C$ is the integration constant corresponding to the black-hole mass of
the Schwarz\-schild--(anti)-de~Sitter solution in the bulk and associated with
what is called ``dark radiation.''

Equation (\ref{1st}) can be solved with respect to the Hubble
parameter:
\begin{equation}\label{first}
H^2+\frac{\kappa}{a^2}=\frac{\rho+\sigma}{3m^2}+\frac2{\ell^2}
\left[1\pm\sqrt{1+\ell^2\left(\frac{\rho+\sigma}{3m^2}-\frac{\Lambda_{\rm
b}}{6} -\frac{C}{a^4}\right)}\right]\, ,
\end{equation}
where $\ell= 2m^2 / M^3$ is the crossover length scale characterizing the
theory. The `$\pm$' signs give the two different branches of braneworld
solutions which, to some extent, are connected with the two different ways in
which the Schwarz\-schild--(anti)-de Sitter bulk space is bounded by the brane:
the inner normal to the brane can point either in the direction of increasing
or decreasing of the bulk radial coordinate. Following \cite{SS}, we refer to
the model with the lower sign (`$-$') as to BRANE1, and to that with the upper
sign (`$+$') as to BRANE2.

In the formal limit of $m \to 0$, the BRANE1 model reduces to the well-known
expression obtained in the Randall--Sundrum case
\begin{equation}\label{rs-limit}
H^2 + \frac{\kappa}{a^2} = \frac{(\rho + \sigma)^2}{9 M^6} + \frac{\Lambda_{\rm
b}}{6} + \frac{C}{a^4} \, ,
\end{equation}
while the BRANE2 model does not exist in this limit. This can be viewed as a
simple consequence of the fact that Eq.~(\ref{1st}), being quadratic with
respect of the Hubble parameter $H$ in the case $m \ne 0$, reduces to a linear
equation with respect to $H$ in the formal limit $m \to 0$.

\section{Static solutions}\label{static}

We proceed to the investigation of static solutions in the braneworld theory.
First, we review the situation in the general relativity theory, where the
static Einstein's universe is known to be unstable.  We apply the same method
for the investigation of the stability of static solutions in the braneworld
theory and demonstrate new possibilities arising in this theory. Firstly,
unlike general relativity, the braneworld theory admits static solutions even
in the case of spatially flat or open universe, which, in the context of the
Randall--Sundrum model, was also noted in \cite{GM}. In the Randall--Sundrum
model, static spatially flat or open universes with matter satisfying the
strong energy condition are possible only with negative value of the
dark-radiation constant $C$, while, in the presence of the scalar-curvature
term in the action for the brane, a spatially flat or open universe can be
static even when this constant is zero. Secondly, these solutions can be stable
in certain region of parameters.

\subsection{The case of general relativity}

In this section, we review the situation in the general relativity theory with
the standard action
\begin{equation}
S_{\rm GR}=-\frac{1}{16\pi G}\int(R-2\Lambda)\sqrt{-g}d^4x+\int L_{\rm
matter}\sqrt{-g}d^4x
\end{equation}
and describe our method for investigation of the stability of
static solutions on the example of this theory. A homogenous and
isotropic universe with density $\rho$ and pressure $p=w\rho$ is
described by the Friedmann equations
\begin{equation}\label{gr-first}
H^2+\frac{\kappa}{a^2}=\frac{8\pi G\rho}3+\frac{\Lambda}3\, ,
\end{equation}
\begin{equation} \label{gr-second}
\frac{\ddot{a}}{a}=-\frac{4\pi G\rho}3(1+3w)+\frac{\Lambda}3\, ,
\end{equation}
giving the following conditions of static solution:
\begin{equation}
\frac{\kappa}{a_0^2}=\frac{8\pi G\rho_0}3+\frac{\Lambda}3\, ,
\end{equation}
\begin{equation}
\frac{4\pi G\rho_0}3(1 + 3w) - \frac{\Lambda}3 = 0\, .
\end{equation}
Introducing small homogeneous deviations from stationarity,
\begin{equation}
a=a_0(1-\epsilon)\, , \quad \epsilon = \epsilon(t) \ll 1 \, ,
\end{equation}
and linearizing the Friedmann equations (\ref{gr-first}) and
(\ref{gr-second}) with respect to the small quantity $\epsilon
(t)$, we have:
\begin{equation}
\frac{2\kappa\epsilon}{a_0^2} = \frac{8\pi G(\rho-\rho_0)}3\, ,
\end{equation}
\begin{equation}\label{gr}
\ddot{\epsilon} = W \epsilon, \quad W=\frac{\kappa(1+3w)}{a_0^2}\, .
\end{equation}
We note that the weak energy condition and casuality requirements restrict the
value of $w$ to lie in the interval $w \in [-1,+1]$ (see, e.g., \cite{HE}) and,
as a result, $\kappa=+1$. Taking this into account, one can see from (\ref{gr})
that the static solution in this theory is unstable for $w \in
\left[-\frac13,1\right]$, and stable for $w \in \left(-1,-\frac13\right)$.

\subsection{The case of braneworld theory}

One can use the method outlined in the previous section to investigate the
braneworld theory. In this section, we find (analytically and graphically) the
regions of existence of static solutions in the parameter space and then
further restrict the space of parameters by the requirement that the described
theory be physically compatible with the observations.

\subsubsection{Static solutions and their perturbation}

In the static case, equations (\ref{second}) and (\ref{1st}) give
\begin{equation}\label{second0}
9M^{6}\left(\frac{\kappa}{a^2}-\frac{\Lambda_{\rm b}}3\right)
+\left[\rho(1+3w)-2\sigma\right] \left(\rho+\sigma -
3m^2\frac{\kappa}{a^2}\right)=0
\end{equation}
and
\begin{equation}\label{1st0}
m^4\left(\frac{\kappa}{a^2}-\frac{\rho+\sigma}{3m^2}\right)^2=
M^{6}\left(\frac{\kappa}{a^2}-\frac{\Lambda_{\rm b}}{6}-\frac{C}{a^4}\right)\,
,
\end{equation}
respectively.  One should note that Eq.~(\ref{1st0}) cannot be obtained by
integrating Eq.~(\ref{second}) in the static case because, before integration,
Eq.~(\ref{second}) is multiplied by $\dot{a}$, which, in this case, is
identically zero. Nevertheless, it can be obtained from the original equations
on the brane (\ref{eq_mot2}) considering the embedding of the brane in the
bulk.  In fact, this equation determines the value of the black-hole constant
$C$ for which a solution of (\ref{second0}) is embeddable in the
Schwarz\-schild--(anti)-de~Sitter bulk.

Looking for static solutions and investigating their stability, we proceed as
in the previous case of general relativity. First, we solve Eq.~(\ref{second})
with respect to $\ddot a / a$ for $\dot a = 0$:
\begin{equation}\label{1st1}
\frac{\ddot{a}}{a}=-\frac{9M^{6}\left(\frac{\kappa}{a^2}-\frac{\Lambda_{\rm
b}}3\right)+ \left[\rho(1+3w)-2\sigma\right] \left(\rho+\sigma -
3m^2\frac{\kappa}{a^2}\right)}{9M^{6}+6m^2\left(\rho+
\sigma-3m^2\frac{\kappa}{a^2}\right)}\, .
\end{equation}
Studying homogeneous perturbations of this solution, we set $\dot{a}=0$ as the
initial condition. If a solution is stable (unstable) with respect to such
perturbations, it will also be stable (unstable) with respect to more generic
perturbations with $\dot{a}\neq 0$. Thus, we assume that the relation between
$\rho$ and $a$ is given by Eq.~(\ref{1st0}) at the initial moment of time.

We express the energy density $\rho$ through $a$ using Eq.~(\ref{1st0}),
\begin{equation}
\rho=-\sigma+3m^2\frac{\kappa}{a^2}\mp
3M^3\sqrt{\frac{\kappa}{a^2}-\frac{\Lambda_{\rm b}}{6}-\frac{C}{a^4}}\, ,
\end{equation}
and substitute it to (\ref{1st1}):
\begin{equation}\label{1st2}
\frac{\ddot{a}}{a}=\frac{A_1}{9M^{6}\mp
18m^2M^3\sqrt{\frac{\kappa}{a^2}-\frac{\Lambda_{\rm b}}{6}-\frac{C}{a^4}}}\, ,
\end{equation}
where
\begin{equation}
A_1 = -9M^6 \left( \frac{\kappa}{a^2} - \frac{\Lambda_{\rm b}}3 \right) \pm
3M^3 A_2 \sqrt{\frac{\kappa}{a^2} - \frac{\Lambda_{\rm b}}{6} -
\frac{C}{a^4}}\, ,
\end{equation}
\begin{equation}
A_2=-3\sigma (1+w)+3(1+3w)m^2\frac{\kappa}{a^2}\mp
3(1+3w)M^3\sqrt{\frac{\kappa}{a^2}-\frac{\Lambda_{\rm b}}{6}-\frac{C}{a^4}}\, .
\end{equation}

After that, we introduce the perturbation of the static solution
\begin{equation}
a=a_0(1-\epsilon)\, , \quad \epsilon (t) \ll 1 \, ,
\end{equation}
where $a_0$ is a stationary solution of (\ref{1st2}), and expand
Eq.~(\ref{1st2}) around the stationary point:
\begin{equation}
\ddot \epsilon = W \epsilon \, ,
\end{equation}
where
\begin{equation}\label{w}
W = \frac{B_1}{1 \pm \frac{2m^2}{M^3} \sqrt{\frac{\kappa}{a_0^2} -
\frac{\Lambda_{\rm b}}{6} - \frac{C}{a_0^4}}}\, ,
\end{equation}
\begin{equation}\begin{array}{l}
\displaystyle B_1 = \frac{2\kappa}{a_0^2}+(1+3w)
\left(\frac{2\kappa}{a_0^2}-\frac{4C}{a_0^4}\right)\mp B_2\pm B_3 \pm B_4\, ,
\smallskip \\
\displaystyle B_2=\frac{\sigma(1+w)}{M^3
\sqrt{\frac{\kappa}{a_0^2}-\frac{\Lambda_{\rm
b}}{6}-\frac{C}{a_0^4}}}\left(\frac{\kappa}{a_0^2} -\frac{2C}{a_0^4}\right)\, ,
\smallskip \\
\displaystyle B_3=\frac{m^2}{M^3}\frac{2\kappa}
{a_0^2}(1+3w)\sqrt{\frac{\kappa}{a_0^2}-\frac{\Lambda_{\rm
b}}{6}-\frac{C}{a_0^4}}\, ,
\smallskip \\
\displaystyle B_4=\frac{m^2}{M^3}\frac{\kappa}{a_0^2}(1+3w)\frac{\frac{\kappa}
{a_0^2}-\frac{2C}{a_0^4}}{\sqrt{\frac{\kappa}{a_0^2}- \frac{\Lambda_{\rm
b}}{6}-\frac{C}{a_0^4}}}\, .
\end{array}
\end{equation}
The lower and upper signs in these equations refer to the BRANE1 and BRANE2
models, respectively.

It can also be noted \cite{SS} that, unlike the general relativity theory, the
braneworld theory allows for empty ($\rho = p = 0$) static solutions.  The
radius (scale factor) of such a universe is given by the expression \cite{SS}
\begin{equation}
a^2 = \frac{\kappa}{\Lambda_{\rm RS}} \left( \frac32 + \frac{\sigma m^2}{M^6}
\right) \, .
\end{equation}
The spatially flat case requires $\Lambda_{\rm RS} = 0$.

\subsubsection{Physical restrictions on the constants of the theory }

Observations indicate that, at the present cosmological epoch, our universe
expands with acceleration, so that the simplest asymptotic condition for such a
universe is that the energy density will decrease monotonically to zero. Thus,
we can consider the low-energy behavior of the theory as $\rho\rightarrow 0$:
\begin{equation}\label{low-energy}
H^2+\frac{\kappa}{a^2}=C_1 + C_2\rho\, , \quad \qquad C_1
> 0\, , \quad C_2 > 0\, ,
\end{equation}
which is valid under the condition
\begin{equation} \label{regime}
\frac{\ell^2 \rho}{3 m^2} \ll 1+\ell^2 \left( \frac{\sigma}{3m^2} -
\frac{\Lambda_{\rm b}}{6} \right) \, ,
\end{equation}
and where we have assumed that the dark-radiation term (with the constant $C$
in (\ref{first})) has negligible contribution.  Assuming regime
(\ref{low-energy}) is realized today, we can write the following minimal
physical restrictions on the constants of the theory:

\bigskip

{\bf BRANE1:} The inequalities
\begin{equation}
C_1 \equiv \frac{\sigma}{3m^2} + \frac2{\ell^2} \left[ 1  - \sqrt{1+\ell^2
\left( \frac{\sigma}{3m^2} - \frac{\Lambda_{\rm b}}{6} \right)} \right]
> 0
\end{equation}
and
\begin{equation}
C_2 \equiv \frac{1}{3m^2} \left[ 1 - \frac{1}{\sqrt{1+\ell^2
\left( \frac{\sigma}{3m^2} - \frac{\Lambda_{\rm b}}{6} \right)}}
\right] > 0
\end{equation}
give
\begin{equation}\label{rest1}
\sigma > 0 \, , \quad \sigma > \frac{\Lambda_{\rm b}m^2}2 \, , \quad
\Lambda_{\rm RS} \equiv \frac{\Lambda_{\rm b}}2 + \frac{\sigma^2}{3 M^6} > 0 \,
.
\end{equation}

\bigskip

{\bf BRANE2:} The inequalities
\begin{equation}
C_1 \equiv \frac{\sigma}{3m^2} + \frac2{\ell^2} \left[ 1 + \sqrt{1+\ell^2
\left(\frac{\sigma}{3m^2} - \frac{\Lambda_{\rm b}}{6}\right)} \right] > 0
\end{equation}
and
\begin{equation}
C_2 \equiv \frac{1}{3m^2} \left[ 1 + \frac{1}{\sqrt{1+\ell^2 \left(
\frac{\sigma}{3m^2} - \frac{\Lambda_{\rm b}}{6} \right)}} \right] > 0
\end{equation}
imply
\begin{equation}\label{rest2}
\sigma > -\frac{6m^2}{\ell^2}\, , \quad \sigma > \frac{\Lambda_{\rm b} m^2}2 -
\frac{3m^2}{\ell^2}\, .
\end{equation}

\bigskip

From these relations, one can see that BRANE2 model with the asymptotic
conditions (\ref{low-energy}), in principle, can have negative brane tension,
while the tension of the BRANE1 model is restricted to  be positive.

The physical restrictions formulated above are necessary for a reasonable
braneworld theory with the asymptotic limit (\ref{low-energy}) reached today,
but they are certainly far from being sufficient for the consistency of a
braneworld model with all cosmological observations (cosmic microwave
background, big-bang nucleosynthesis, etc).  This important issue, which, in
fact, constitutes the main subject of investigations in the braneworld theory,
clearly lies beyond the scope of the present paper and will be considered in
our future publications.  At the moment, one can only note that the values of
the constants $C_1$ and $C_2$ in a braneworld that today evolves in regime
(\ref{low-energy}) are roughly constrained by the values of the detected
cosmological constant and gravitational coupling, respectively, as
\begin{equation}
C_1 \simeq 0.7 H_0^2 \, , \qquad C_2 \simeq \frac{8 \pi G_{\rm N}}{3} \, ,
\end{equation}
where $H_0 \approx 73\, {\rm km/sec} \cdot {\rm Mpc}$ is the current value of
the Hubble parameter, and $G_{\rm N}$ is Newton's constant.  With these
constraints, there remains a considerable freedom in the parameter space of the
theory, which is to be further restricted by a more careful analysis.

If condition (\ref{regime}) is not satisfied today, then regime
(\ref{low-energy}) has not yet been reached, and one should turn to the general
evolution equation (\ref{first}), which has a rather rich and interesting
phenomenology for modeling dark energy \cite{SS,AS}. For example, the current
acceleration of the universe expansion may be described by a braneworld model
which has $C_1 = 0$ in the asymptotic expansion (\ref{low-energy}). According
to this model, the current acceleration of the cosmological expansion is
temporary, and the future evolution proceeds in a matter-dominated regime (see
\cite{SS} for more details).  Another interesting example is the model of
loitering braneworld \cite{loiter}, which can be realised even in the spatially
flat universe.  It is the negative dark radiation (the constant $C$ in
Eq.~(\ref{first})) which plays the crucial role in the dynamics of this model
and which, therefore, also needs to be constrained by cosmological
observations.

\subsubsection{Stability of the spatially flat static solutions}

For $\kappa=0$, our equations (\ref{second0}), (\ref{1st0}) and (\ref{w})
simplify to
\begin{equation}
{} -3M^{6}\Lambda_{\rm b} + \left[ \rho (1+3w) - 2 \sigma \right] \cdot
\left(\rho + \sigma \right)=0\, ,
\end{equation}
\begin{equation}
(\rho + \sigma)^2 = {} - 9M^{6} \left( \frac{\Lambda_{\rm b}}{6} +
\frac{C}{a^4} \right)
\end{equation}
and
\begin{equation}
W=\frac{1}{1+\frac{2m^2}{3M^{6}}(\rho_0+\sigma)}\left[-(1+3w)
\frac{4C}{a_0^4}+\frac{3\sigma(1+w)}{\rho_0+\sigma} \frac{2C}{a_0^4}\right]\, ,
\end{equation}
respectively.

Introducing the dimensionless parameters
\begin{equation}
x = \frac{\rho_0\ell^2}{6m^2 }= \frac{2m^2\rho_0}{3M^{6}} \, , \quad y =
\frac{\sigma \ell^2}{6m^2} = \frac{2m^2\sigma}{3M^{6}} \, , \quad z =
\frac{\Lambda_{\rm b}\ell^2}{6} = \frac{2m^4\Lambda_{\rm b}}{3M^{6}}\, ,
\end{equation}
we can express the quantities that we used before in their terms. Since only
two of these parameters are independent, it is convenient to express the
relevant quantities in terms of $x$ and $y$. We have
\begin{equation}
z=\frac{1}2(x+y)\left[(1 + 3w)x - 2y \right]\, ,\label{z}
\end{equation}
\begin{equation}\label{C}
\frac{C}{a_0^4}=-\frac{3M^{6}}{8m^4} (1 + w) x(x + y)\, ,
\end{equation}
\begin{equation}\label{W}
W=\frac{3M^{6}}{4m^4}(1+w)x\frac{2(1+3w)x - (1-3w)y}{1+x+y}\, .
\end{equation}

Taking into account the physical restrictions (\ref{rest1}) and (\ref{rest2})
and equations (\ref{z}) and (\ref{W}), we can indicate the domains of existence
of static solutions (both stable and unstable) in the $(x,y)$ plane.

The natural condition $\rho_0 > 0$ leads to the restriction $x > 0$.

For BRANE1, restrictions (\ref{rest1}) additionally give
\begin{equation}
y > 0\, , \quad 4y > (x + y) \left[ (1 + 3w) x - 2 y \right] \, , \quad (1 +
3w) x > (1 - 3w) y \, .
\end{equation}

For BRANE2, restrictions (\ref{rest2}) additionally give
\begin{equation}
y > -1 \, , \quad 4y + 2 > (x + y) \left[ (1 + 3w) x - 2y \right] \, .
\end{equation}

The corresponding domains are shown below in Figs.~\ref{f:1} and \ref{f:2} for
the two important cases $w = 0$ and $w = 1/3$. The shaded regions in these
figures correspond to unstable static solutions ($W > 0$), and the dotted
regions correspond to stable static solutions ($W \le 0$).  In general, stable
static solutions exist only for the BRANE2 model in the case $w < 1/3$.

\begin{figure}[tbh!]
\centerline{
\psfig{figure=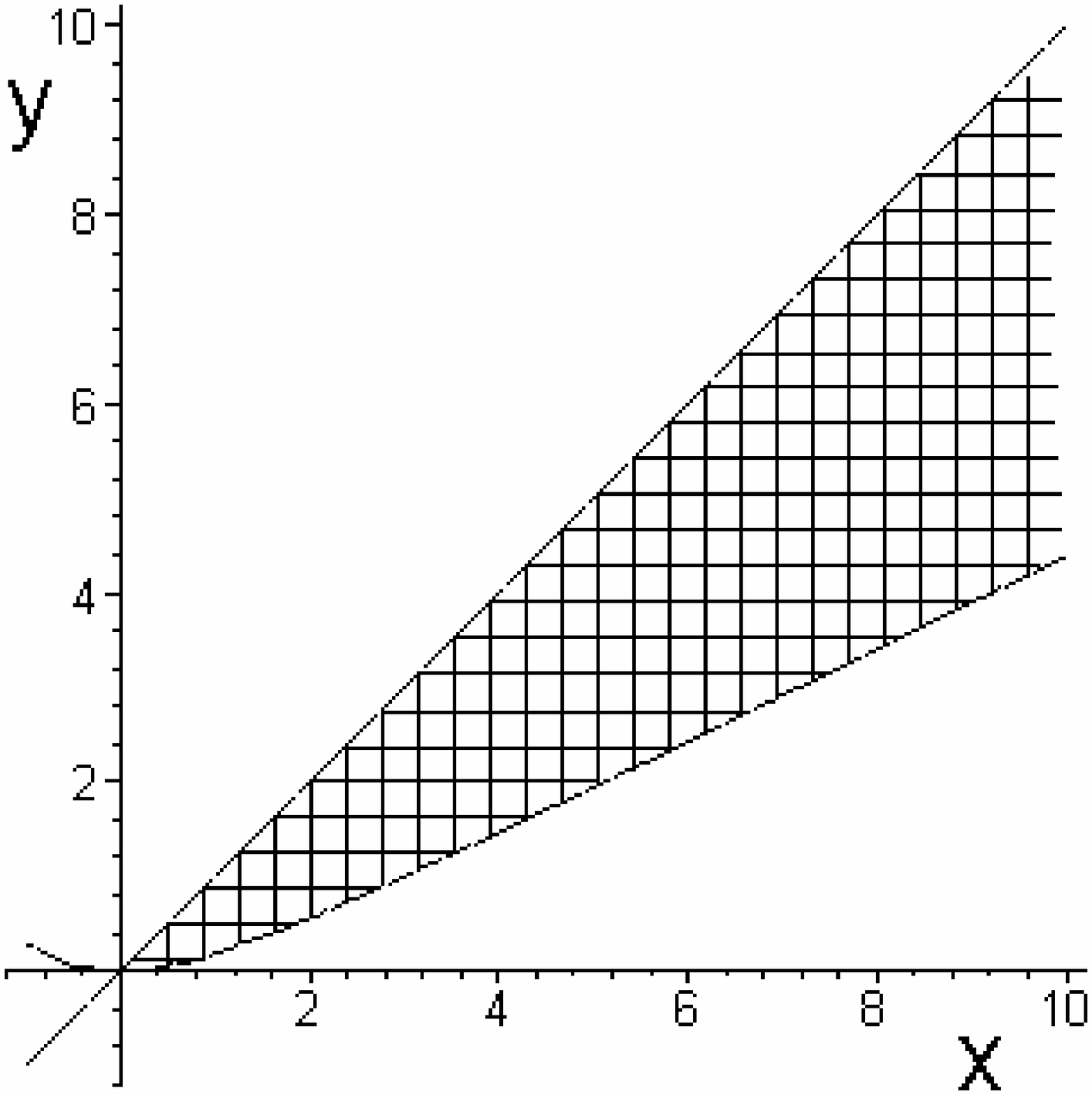,width=0.55\textwidth,angle=90}
\psfig{figure=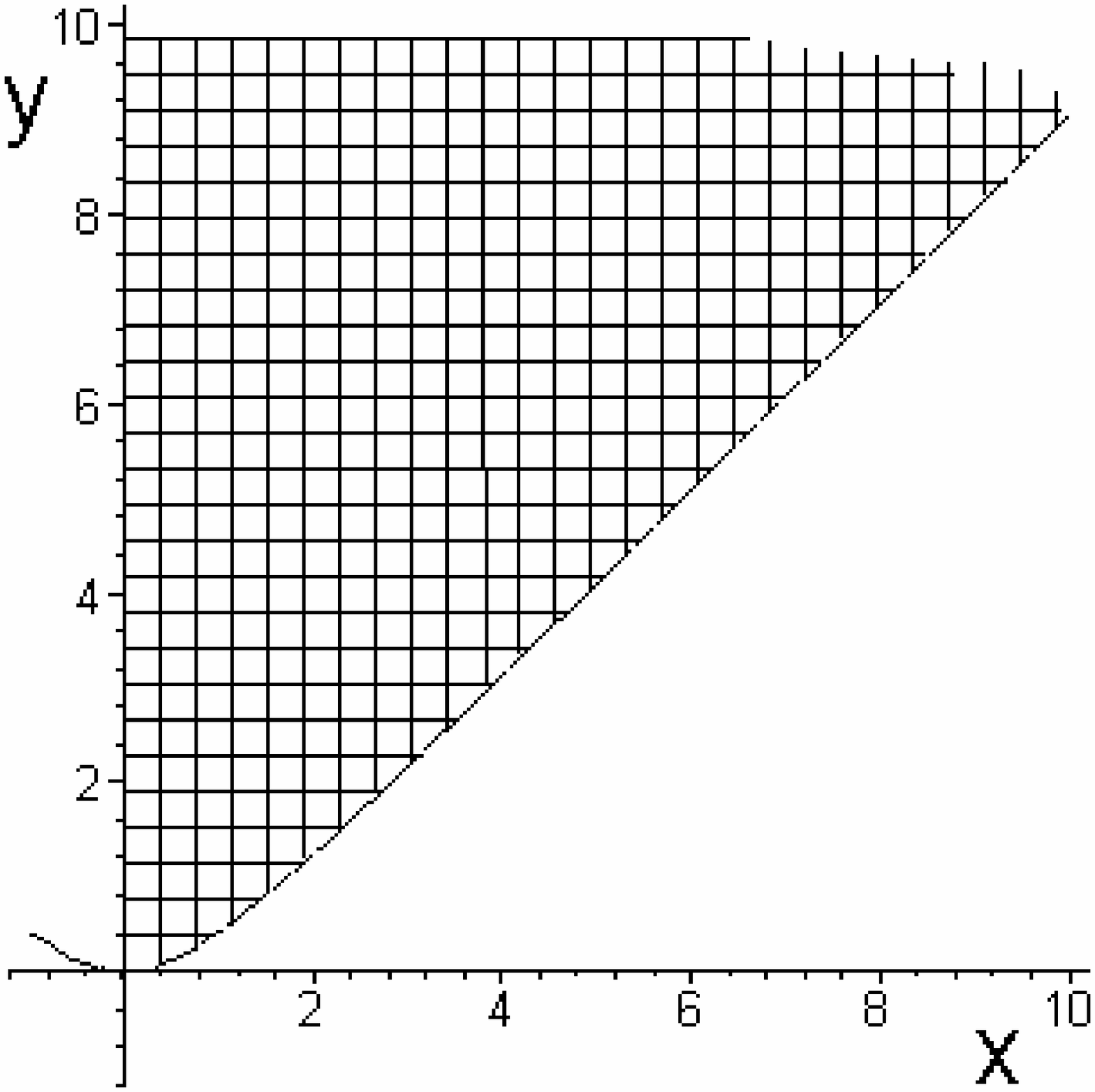,width=0.55\textwidth,angle=90} }
\caption{\small BRANE1, $w=0$ (left) and $w = 1/3$ (right). The
shaded regions indicate the physically allowable parameters,
corresponding to unstable static solutions.} \label{f:1}
\vspace{1cm}
\end{figure}

\begin{figure}[tbh!]
\centerline{
\psfig{figure=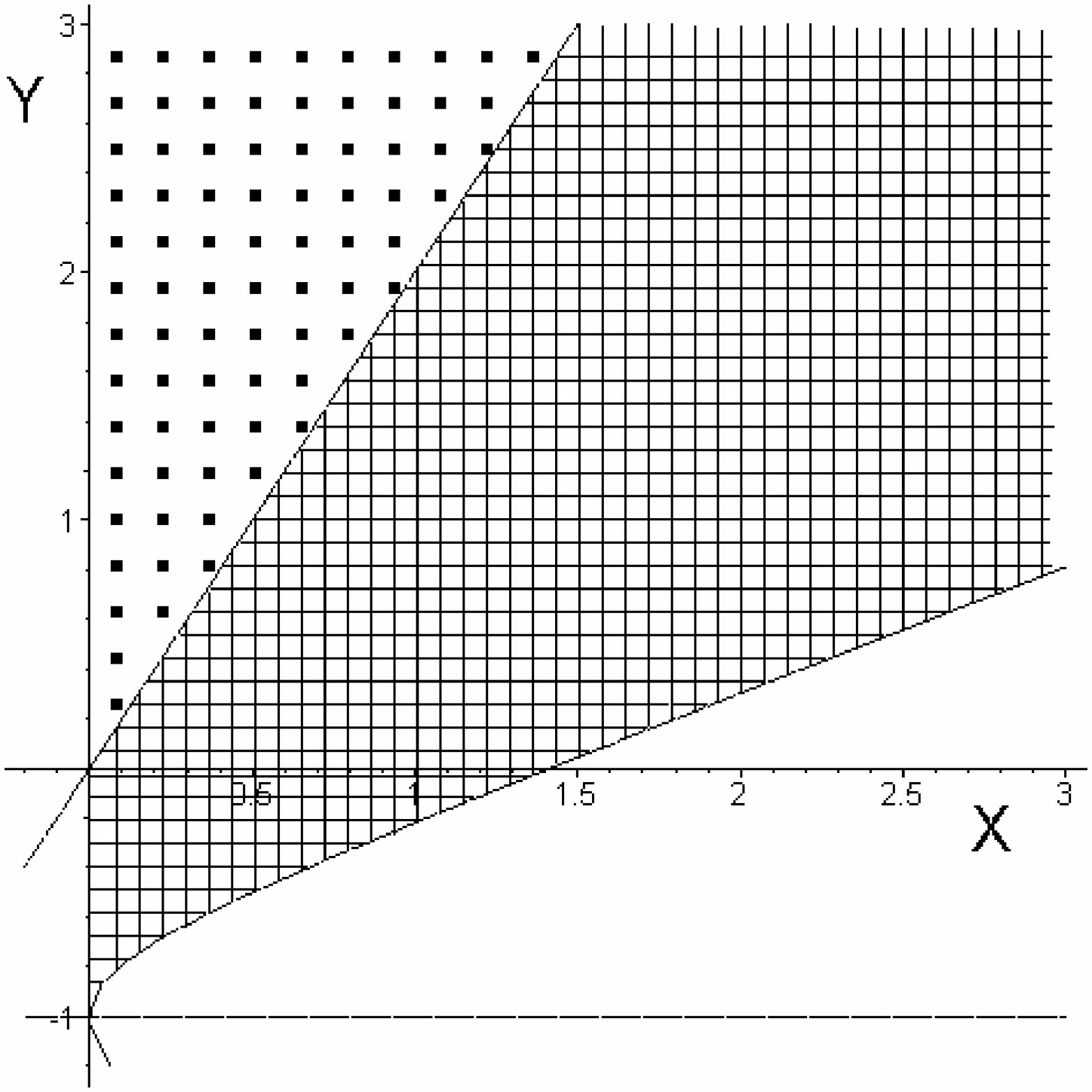,width=0.55\textwidth,angle=90}
\psfig{figure=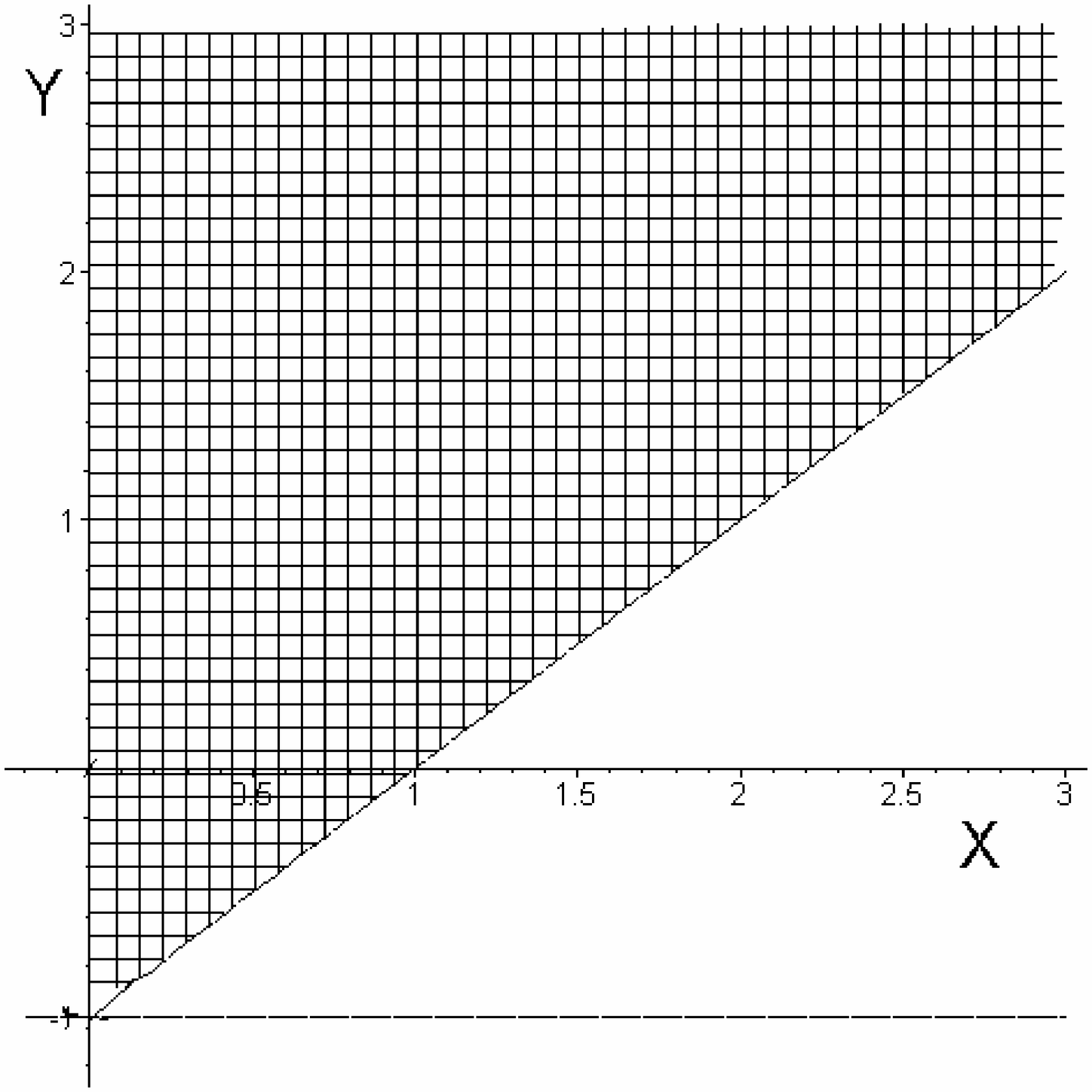,width=0.55\textwidth,angle=90} }
\caption{\small BRANE2, $w=0$ (left) and $w = 1/3$ (right). The
shaded regions of the physically allowable parameters correspond
to unstable static solutions, and the dotted region corresponds to
stable static solutions.} \label{f:2}
\end{figure}

It also follows from (\ref{C}) that spatially flat static
solutions with matter satisfying the strong energy condition can
exist in the BRANE2 model with $C = 0$, i.e., with zero
dark-radiation term, in contrast with the Randall--Sundrum model,
where spatially flat or open static solutions require negative
dark-radiation term. This difference is connected with the fact
that the physically admissible range of parameters in the
Randall--Sundrum model requires positive brane tension while, in
the presence of the scalar-curvature term in the action for the
brane, this restriction is weakened, and negative brane tensions
are also allowed. As follows from (\ref{C}), the condition $C = 0$
is equivalent to $x + y = 0$, and the region with negative $y$ in
the admissible domain of parameters allowing for this condition
can be seen in Fig.~\ref{f:2}.

Stability conditions for a spatially non-flat universe ($\kappa \ne 0$) are
further considered by a different method in Sec.~\ref{static-case}.

\section{Non-static solutions}\label{nonstatic}

In this section, we describe non-static solutions of the braneworld theory
under investigation and also determine their stability.  To do this, we use a
set of convenient phase-space variables similar to those introduced in
\cite{CS,GE}. The critical points of the system of differential equations in
the space of these variables describe interesting non-static solutions. A
method for evaluating the eigenvalues of the critical points of the Friedmann
and Bianchi models was introduced by Goliath and Ellis \cite{GE} and further
used in the analysis by Campos and Sopuerta \cite{CS} of the Randall--Sundrum
braneworld theory. In this section, we extend their investigation to the case
where the scalar-curvature term is also present in the action for the brane,
but we restrict ourselves to the homogeneous and isotropic cosmology.

\subsection{ The case of the Randall--Sundrum braneworld model}\label{method}

In this subsection, we briefly describe the method and results of \cite{CS},
where the model with $m=0$ was under consideration.  After setting $m = 0$,
Eqs. (\ref{second}) and (\ref{first}) turn to
\begin{equation}\label{m2}
9M^{6}\left[\frac{\ddot{a}}{a}+H^2+\frac{\kappa}{a^2}-\frac{\Lambda_{\rm
b}}3\right] +[\rho(1+3w)-2\sigma](\rho+\sigma)=0
\end{equation}
and
\begin{equation}\label{m1}
H^2+\frac{\kappa}{a^2} = \frac{(\rho+\sigma)^2}{9M^{6}} + \frac{C}{a^4} +
\frac{\Lambda_{\rm b}}{6}\, ,
\end{equation}
respectively.  Here, for simplicity, we consider the case $C > 0$. The last
equation can be written as follows:
\begin{equation}
H^2 = \frac{2\sigma}{9M^{6}} \rho \left(1 + \frac{\rho}{2\sigma} \right) -
\frac{\kappa}{a^2} + \frac{C}{a^4} + \frac13 \left( \frac{\Lambda_{\rm
b}}2+\frac{\sigma^2}{3M^{6}}\right) \, .
\end{equation}

Using the definition of the four-dimensional cosmological constant
\begin{equation}
\Lambda_{\rm RS} = \frac{\Lambda_{\rm b}}2+\frac{\sigma^2}{3M^{6}}\, ,
\end{equation}
which is proportional to the Randall--Sundrum constraint \cite{RS}, we
introduce the notation similar to those of \cite{CS}:
\begin{equation}
\Omega_{\rho}=\frac{2\sigma\rho}{9M^{6}H^2}\, , \quad
\Omega_{k}=-\frac{\kappa}{a^2H^2}\, , \quad \Omega_{\Lambda}=\frac{\Lambda_{\rm
RS}}{3H^2}\, , \quad \Omega_{\lambda}=\frac{\rho^2}{9M^{6}H^2}\, , \quad
\Omega_C = \frac{C}{a^4 H^2} \, .
\end{equation}

The authors of \cite{CS} work in the five-dimensional $\Omega$-space
$(\Omega_{\rho},\Omega_{k},\Omega_{\Lambda},\Omega_{\lambda},\Omega_C)$.
However, the $\Omega$ parameters are not all independent since they are related
by the condition
\begin{equation}\label{constraint}
\Omega_{\rho}+\Omega_{k}+\Omega_{\Lambda}+\Omega_{\lambda} + \Omega_C =1\, .
\end{equation}
Taking this constraint into account leaves us with a four-dimensional space and
only four eigenvalues of the static solution to be found.

Introducing the primed time derivative
$$
' = \frac{1}{H} \frac{d}{dt} \, ,
$$
one obtains the system of first-order differential equations \cite{CS}
\begin{equation}\begin{array}{l}
\Omega_{\rho}'=[2(1+q)-3(1+w)]\Omega_{\rho}\, , \smallskip \\
\Omega_{k}'=2q\Omega_{k}\, , \smallskip \\
\Omega_{\Lambda}'=2(1+q)\Omega_{\Lambda}\, , \smallskip \\
\Omega_{\lambda}'=2[(1+q)-3(1+w)]\Omega_{\lambda}\, , \smallskip \\
\Omega_C' = 2 (q - 1) \Omega_C \, ,
\end{array}
\end{equation}
where
\begin{equation}
q = \frac{1 + 3w}{2} \Omega_{\rho} - \Omega_{\Lambda} + (3w + 2)
\Omega_{\lambda} + \Omega_C\, .
\end{equation}
The behavior of this system of equations in the neighborhood of its stationary
point is determined by the corresponding matrix of its linearization. The real
parts of its eigenvalues tell us whether the corresponding cosmological
solution is stable or unstable with respect to the homogeneous perturbations.

As was said above, only four parameters $\Omega$ are independent because of
constraint (\ref{constraint}).  In each of the cases listed below, we choose
independent parameters conveniently, and our choice is reflected in the indices
of the corresponding eigenvalues.

\bigskip

\noindent {\bf (1)} \ The Friedmann case, or
$(\Omega_{\rho},\Omega_{k},\Omega_{\Lambda},\Omega_{\lambda},\Omega_C) =
(1,0,0,0,0)$. We have
\begin{equation}
q=\frac{1+3w}2\, ,
\end{equation}
and the eigenvalues are
\begin{equation}
\lambda_{k}=1+3w\, , \quad \lambda_{\Lambda}=3(1+w)\, , \quad
\lambda_{\lambda}=-3(1+w)\, , \quad \lambda_C = 3 w - 1 \, .
\end{equation}

\bigskip

\noindent {\bf (2)} \ The Milne case, or $(\Omega_{\rho}, \Omega_{k},
\Omega_{\Lambda}, \Omega_{\lambda},\Omega_C) = (0,1,0,0,0)$. We have
\begin{equation}
q=0\, ,
\end{equation}
and the eigenvalues are
\begin{equation}
\lambda_{\rho}=-(1+3w)\, , \quad \lambda_{\Lambda}=2\, , \quad
\lambda_{\lambda}=-2(2+3w)\, , \quad \lambda_C = - 2 \, .
\end{equation}

\bigskip

\noindent {\bf (3)} \ The de~Sitter case, or $(\Omega_{\rho}, \Omega_{k},
\Omega_{\Lambda}, \Omega_{\lambda}, \Omega_C) = (0,0,1,0,0)$. We have
\begin{equation}
q=-1\, ,
\end{equation}
and the eigenvalues are
\begin{equation}
\lambda_{\rho}=-3(1+w)\, , \quad \lambda_{k}=-2\, , \quad
\lambda_{\lambda}=-6(1+w)\, , \quad \lambda_C = - 4 \, .
\end{equation}

\bigskip

\noindent {\bf (4)} \ The case considered in \cite{BDEL}, or $(\Omega_{\rho},
\Omega_{k}, \Omega_{\Lambda}, \Omega_{\lambda}, \Omega_C) = (0,0,0,1,0)$. We
have
\begin{equation}
q=2+3w\, ,
\end{equation}
and the eigenvalues are
\begin{equation}
\lambda_{\rho}=3(1+w)\, , \quad \lambda_{k}=2(2+3w)\, , \quad
\lambda_{\Lambda}=6(1+w)\, , \quad \lambda_C = 2 (1 + 3w) \, .
\end{equation}

\bigskip

\noindent {\bf (5)} \ The dark-radiation case, or $(\Omega_{\rho}, \Omega_{k},
\Omega_{\Lambda}, \Omega_{\lambda}, \Omega_C) = (0,0,0,0,1)$. We have
\begin{equation}
q = 1\, ,
\end{equation}
and the eigenvalues are
\begin{equation}
\lambda_{\rho} = 1 - 3w \, , \quad \lambda_{k} = 2\, , \quad \lambda_\Lambda =
4 \, , \quad \lambda_{\lambda} = - 2(1 + 3w) \, .
\end{equation}

All these results for the case $m = 0$ were obtained in \cite{CS}.

\subsection{The case $m\neq 0$}

In this subsection, we apply the method of \cite{CS} to the case of braneworld
theory with the scalar-curvature term in the action for the brane, i.e., where
$m \ne 0$. We introduce the notation
\begin{equation}\begin{array}{l}
\displaystyle B=\frac{\Lambda_{\rm b}}{3 \ell^2 H^4}\, , \quad P
=\frac{\rho}{6m^2H^2}\, ,
\quad S=\frac{\sigma}{6m^2H^2}\, , \bigskip \\
\displaystyle L=\frac{1}{\ell^2H^2}\, , \quad K=\frac{\kappa}{a^2H^2}\, , \quad
D=\frac{C}{\ell^2 a^4H^4}\, ,
\end{array}
\end{equation}
and express the deceleration parameter
\begin{equation}
q=-\frac{\ddot{a}}{H^2a}
\end{equation}
from the second-order differential equation (\ref{second}):
\begin{equation}\label{acc}
q = \frac{L(1+K) -B +[P(1+3w)-2S] [P + S-
\frac{1}2(1+K)]}{L+P+S-\frac{1}2(1+K)}\, .
\end{equation}

Equation (\ref{1st}) implies the constraint
\begin{equation} \label{constraint1}
\left[ 1 + K - 2 (P + S) \right]^2 = 4 L \left( 1 + K \right) - 2 B - D \, .
\end{equation}

Note that the introduced variables are infinite for $H=0$ and $m=0$ and,
therefore, cannot describe these particular cases. The second case was already
considered in the previous subsection and in \cite{CS}, and the first case will
be considered below in Sec.~\ref{static-case} (using different variables).

Introducing the prime derivative
$$
'=\frac{1}{H}\frac{d}{dt}\, ,
$$
we obtain the system of equations describing the evolution of our dimensionless
parameters:
\begin{equation}\label{system}
\begin{array}{l}
P'=-P(1+3w-2q)\, , \smallskip \\
S'=2S(1+q)\, , \smallskip \\
L'=2L(1+q)\, , \smallskip \\
B'=4B(1+q)\, , \smallskip \\
K'=2Kq\, , \smallskip \\
D'=4Dq \, ,
\end{array}
\end{equation}
where $q$ is given by Eq.~(\ref{acc}).  The critical points of this system are
those values of $(P$, $S$, $L$, $B$, $K$, $D)$ which nullify its right-hand
side. Since three equations in this system have similar form, there are three
different options for the generic values of $w$, and there are also three
special cases, namely, $w = -1$, $w =  1/3$ and $w = - 1/3$, which we consider
separately.

\subsubsection{Generic situation}\label{generic}

\hspace{\parindent} {\bf Case 1.\,} $1 + 3w - 2q = 0$ and $S = L = B = K = D =
0$. Using (\ref{constraint1}), we obtain
\begin{equation}
P = \frac{1}2\, .
\end{equation}

Hence,
\begin{equation}
H^2 = \frac{\rho}{3 m^2} \, ,
\end{equation}
which describes the standard Friedmann universe and is similar to
case~1 of the previous subsection.

\bigskip

{\bf Case 2.\,} $q = 0$ and $P = S = L = B = 0$.

In this case, Eq.~(\ref{acc}) gives the identity, and the constraint equation
(\ref{constraint1}) reads
\begin{equation}
(1 + K)^2 = - D \, .
\end{equation}
The universe described by this solution is dominated by the spatial curvature
and/or dark radiation with negative constant $C$. In the absence of dark
radiation, we have the spatially open Milne universe, similarly to case~2 of
the previous subsection.

\bigskip

{\bf Case 3.\,} $q = - 1$ and $P = K = D = 0$. Using (\ref{acc}) or
(\ref{constraint1}), we obtain
\begin{equation}
-B=\frac{1}2+2S^2-2S-2L \, .
\end{equation}
This gives us only one equation for the three parameters $L$, $B$, and $S$.
This condition can also be obtained from the first-order differential equation
(\ref{1st}). If we wish to specify the theory (BRANE1 or BRANE2) to which our
solution belongs, we can use the equation following directly from
(\ref{first}):
\begin{equation}
\frac12 - S - L = \pm \sqrt{L^2 + 2LS - \frac12 B} \, .
\end{equation}

All solutions have $H={\rm const}$ (the de~Sitter case). Moreover, it follows
from (\ref{dS}) below that all such solutions are asymptotically stable for all
realistic $w$.

The Hubble constant $H$ can be obtained from the first-order equation
(\ref{first}) with $C=0$, $\rho=0$, and $\kappa=0$:
\begin{equation}\label{h2}
H^2=\frac{\sigma}{3m^2}+\frac2{\ell^2}
\left[1\pm\sqrt{1+\ell^2\left(\frac{\sigma}{3m^2}-\frac{\Lambda_{\rm
b}}{6}\right)}\right].
\end{equation}

This case is similar to case~3 of the previous subsection.

\bigskip

We see that there are no analogues of the solution \cite{BDEL} (case~4 of the
previous subsection) in the generic model with $m \ne 0$. This can be
understood from the comparison of the high-energy limits of equations
(\ref{first}) and (\ref{rs-limit}).  In the first case, the high-energy
behaviour is dominated by the usual linear term with respect to the energy
density on the right-hand side, with negligible contribution from the part with
the square root. Thus, it reproduces the usual Friedmannian behaviour. In the
second case, corresponding to $m = 0$, the high-energy evolution is dominated
by the term quadratic in the energy density.

As for case~5 of the previous subsection describing a universe dominated by
dark radiation, it is now replaced by case~2, in which the dark-radiation
contribution evolves similarly to that of the spatial curvature.  In the
absence of dark radiation, it reduces to case~2 of the previous subsection,
describing the Milne universe.

\subsubsection{Special cases} \label{special}

Apart from the generic cases, there exist special cases $w = -1$ and $w = -
1/3$, in which there arise an additional similarity in the form of otherwise
different equations in (\ref{system}).

\bigskip

{\bf Case 1.\,} $1 + 3w = -2$, or $w = -1$.

In this case, for $q = -1$, $P$ is not necessarily equal to zero. Matter of
this type is equivalent to the four-dimensional cosmological constant (brane
tension $\sigma$) and can be effectively ``eliminated'' by a simple
redefinition of $\sigma$.

\bigskip

{\bf Case 2.\,} $1+3w=0$, or $w = -1/3$.

This case is a little more complicated than the previous one. Now $P$ is not
necessarily equal to zero for $q = 0$.  The constraint equation
(\ref{constraint1}) in this case yields:
\begin{equation}
\left( 1 + K - \frac12 P \right)^2 = - D \, .
\end{equation}
This solution corresponds to the regime where the matter with negative pressure
$p = - \rho / 3$, the spatial-curvature term and the dark-radiation term (with
negative constant $C$) evolve similarly in the cosmological equation
(\ref{first}), so that one has, asymptotically as $a \to 0$,
\begin{equation}
H^2 \approx - \frac{\kappa}{a^2} + \frac{\rho_0}{3 m^2} \left(\frac{a_0}{a}
\right)^2 \pm \frac{2 \sqrt{- C}}{\ell a^2} \, .
\end{equation}

\subsection{Stability of the generic solutions}

To study the stability of a critical point of our system (\ref{system}), we
must expand the system in the neighborhood of this point and then calculate the
eigenvalues of the corresponding evolution matrix.  We introduce the notation:
\begin{equation}
P = P_0 + r\, , \quad S = S_0 + s\, , \quad B = B_0 + b\, , \quad L = L_0 + \xi
\, ,\quad K = K_0 + k\, , \quad D = D_0 + d \, ,
\end{equation}
and consider the same cases as in Sec.~\ref{generic}.  In doing this, we note
that the constraint equation (\ref{constraint1}) reduces the number of
variables to five, and, moreover, the groups of variables \{$S$, $L$, $B$\} and
\{$K$, $D$\} remain proportional between themselves during the evolution, so
that the corresponding eigenvalues should also be proportional.

\bigskip

{\bf Case 1.\,} $q_0 = \frac{1 + 3w}2$, and we obtain the system
\begin{equation}
s' = 3s(1 + w)\, , \quad \xi' = 3\xi(1 + w)\, , \quad b' = 6b(1 + w)\, , \quad
k' = k(1 + 3w)\, , \quad d' = 2 d(1 + 3w)\,.
\end{equation}
Its eigenvalues are:
\begin{equation}
\lambda_S = \lambda_{L} = 3(1+w)\, , \quad \lambda_B = 6(1 + w) \, , \quad
\lambda_K = 1+3w\, , \quad \lambda_D = 2 (1 + 3w) \, .
\end{equation}

\bigskip

{\bf Case 2.\,} $q_0 = 0$. In this case, the set of stationary points is a
curve in the parameter space; hence, one of its eigenvalues will be zero.  We
have the system of equations for the conveniently chosen variables
\begin{equation}
r' = -r(1 + 3w) \, , \quad  s' = 2s \, , \quad \xi' = 2 \xi \, , \quad b' = 4b
\, , \quad k' = 2 K_0 \, \delta q (r, s, \xi, b) \, ,
\end{equation}
where $\delta q$ is the linearized expression (\ref{acc}) for the deceleration
parameter $q$ (which does not depend on $k$). Its eigenvalues are
\begin{equation}\label{dS}
\lambda_1 = 0 \, , \quad \lambda_{P} = -(1 + 3w) \, , \quad \lambda_S =
\lambda_{L} = 2 \, , \quad \lambda_{B} = 4 \, .
\end{equation}

\bigskip

{\bf Case 3.\,} $q_0 = -1$.  In this case, the set of stationary points is a
two-dimensional hypersurface in the parameter space, so two of the eigenvalues
will be zero. The system of equations is
\begin{equation}\begin{array}{l}
r' = -3r (1 + w)\, \quad k' = - 2 k \, , \quad d' = - 4 d\, , \medskip \\
s' = 2 S_0\, \delta q (r, k, d, s, \xi) \, , \quad \xi' = 2 L_0\, \delta q (r,
k, d, s, \xi) \, .
\end{array}
\end{equation}
Its eigenvalues are:
\begin{equation}
\lambda_1 = \lambda_2 = 0 \, , \quad \lambda_{P} = -3(1 + w) \, , \quad
\lambda_K = -2 \, , \quad \lambda_D = -4\,.
\end{equation}

If some of the eigenvalues are positive, then the corresponding solution is
unstable. Summarizing the results obtained, we conclude that the Friedmann
solution is a repeller (always unstable) for all $w > - 1$, and is a saddle
point for $- 1 < w < - 1/3$; the dark-radiation static solution is a repeller
for $-1 < w < - 1/3$, and is a saddle point for $w > - 1/3$; the de~Sitter
solution is an attractor (always stable) for all $w > -1$.

\section{Static case with nonzero spatial curvature} \label{static-case}

To investigate this case, we define the new appropriate variables by dividing
those of the previous section by $L$. Specifically,
\begin{equation}
\tilde P = \frac{P}{L} = \frac{\rho \ell^2}{6 m^2}\, , \quad \tilde B =
\frac{B}{L}\, , \quad \tilde K = \frac{K}{L}\, , \quad \tilde S = \frac{S}{L}
\, , \quad \tilde D = \frac{D}{L^2} \, ,\quad N = \frac{1}{\sqrt{L}} = \ell H
\, .
\end{equation}
Obviously, these variables are finite in the limit $H\rightarrow 0$. Moreover,
only $\tilde{P}$, $\tilde{K}$, $\tilde{D}$ and $N$ vary with time while $\tilde
B$ and $\tilde S$ are constants. We introduce also the new primed derivative
$$
' = \ell \frac{d}{dt} \, .
$$
Then
\begin{equation}\begin{array}{l}
\displaystyle N' = \ell^2 \dot H = \ell^2 \frac{\ddot a}{a} - N^2 \, ,
\medskip \\
\displaystyle \tilde{P}' = -3N \tilde{P}(1 + w) \, ,
\medskip \\
\displaystyle \tilde{K}' = -2N \tilde{K} \, ,
\medskip \\
\displaystyle \tilde{D}' = -4N \tilde{D} \, .
\end{array}
\end{equation}
Stationarity of solution implies
\begin{equation}
N=0\, , \quad \frac{\ddot{a}}{a}=0\, .
\end{equation}
Since
\begin{equation}
\frac{\ddot{a}}{a} = -\frac{1}{\ell^2}\frac{N + \tilde{K} - \tilde{B} + \left[
\tilde{P}(1 + 3w) -2\tilde{S} \right] \left[ \tilde{P} + \tilde{S} -
\frac{1}2(N + \tilde{K}) \right]}{1 + \tilde{P} + \tilde{S}
 - \frac{1}2(N + \tilde{K})}\, ,
\end{equation}
we obtain the equation
\begin{equation}
\tilde{K} - \tilde{B} + \left[ \tilde{P}(1 + 3w) - 2\tilde{S} \right] \left[
\tilde{P} + \tilde{S} - \frac{1}2\tilde{K} \right] = 0 \, .
\end{equation}
Introducing the perturbed variables
\begin{equation}
N=n \, , \quad \tilde{K} = K_0 + k\, , \quad \tilde{P} = P_0 + r\,
, \quad \tilde{D} = D_0 + d\, , \quad \tilde{S} = S_0\, , \quad
\tilde{B} = B_0 \, ,
\end{equation}
and linearizing the system with respect to $n$, $k$, $d$, and $r$,
we obtain
\begin{equation}\begin{array}{ll}
n' &= \displaystyle {} -\frac{k \left( 1 - \frac{1}2 \left[P_0(1 + 3w) -
2S_0\right] \right) + r\left[2P_0(1 + 3w) - S_0 (1 - 3 w) - \frac{K_0}2 (1 +
3w) \right] } {1 + P_0 + S_0 - \frac{K_0}2} \\ &\equiv Fk
+ Gr \, , \smallskip \\
k' &= {} - 2nK_0 \, , \smallskip \\
r' &= {} - 3nP_0(1 + w) \, ,\smallskip \\
d' &= {} - 4nD_0 \, ,
\end{array}
\end{equation}
or, introducing the vector notation $\vec{z}=(d,k,r,n)$,
\begin{equation}
\vec{z}\,' = X\vec{z} \, ,
\end{equation}
where
\begin{equation}
X=\left(
\begin{array}{cccc}
  0 & 0 & 0 & -4D_0 \\
  0 & 0 & 0 & -2K_0 \\
  0 & 0 & 0 & -3P_0(1+w) \\
  0 & F & G & 0 \\
\end{array}
\right).
\end{equation}
The eigenvalue equation for this system is:
\begin{equation}
- \lambda^4 - 2 \lambda^2 K_0 F - 3 \lambda^2 (1 + w) P_0 G = 0 \,
.
\end{equation}
Therefore,
\begin{equation}
\lambda_{1,\,2} = 0 \, , \quad \lambda_{3,\,4} = \pm \ell \sqrt{W}
\end{equation}
with
\begin{eqnarray}\label{W1}
W &=& \frac{2K_0 \left[1 - \frac{1}2 (P_0 (1+3w) - 2S_0)\right]}{\ell^2 \left(1
+ P_0 + S_0 - \frac{K_0}2 \right)} \nonumber \\ \nonumber \\ &{}& {} + \frac{3
P_0 (1 + w) \left[ 2P_0(1 + 3 w) - S_0 ( 1 - 3 w) - \frac{K_0}2 (1 + 3 w)
\right]} {\ell^2 \left(1 + P_0 + S_0 - \frac{K_0}2 \right)} \, .
\end{eqnarray}
We note that two of the eigenvalues are equal to zero. In principle, the
complicated expression (\ref{W1}) determines the regions of stability ($W \le
0$) and instability ($W > 0$) of the static solutions of the theory.

In the partial case $\kappa=0$ (or $K_0 = 0$) that was considered in
Sec.~\ref{static}, we obtain
\begin{equation}\label{ok1}
W = \frac{3 P_0 (1 + w) \left[ 2 P_0 (1 + 3w) + S_0 (3w - 1) \right]} {\ell^2
\left(1 + P_0 + S_0 \right)} \, .
\end{equation}

In the limit $m \to 0$, our result reproduces the corresponding result of
\cite{CS}. Indeed, we have
\begin{equation}
K_0=\frac{\kappa \ell^2}{a^2}=\frac{4\kappa m^4}{a^2M^{6}}\, , \quad
P_0=\frac{\rho \ell^2}{6m^2}=\frac{4\rho m^2}{6M^{6}}\, , \quad
S_0=\frac{4\sigma m^2}{6M^{6}}\, .
\end{equation}
Then, in the limit $m \to 0$,
\begin{equation}
W=\frac{2\kappa}{a^2}+\frac{\rho(1+w)}{3M^{6}}[2\rho(1+3w)+\sigma(3w-1)]\, .
\end{equation}
Expressing $\kappa / a^2$ through $\rho$ and $\sigma$ from (\ref{m2}) and
(\ref{m1}),
\begin{equation}
\frac{\kappa}{a^2}=\frac{\rho(\rho+\sigma)(1+w)}{3M^{6}}\, ,
\end{equation}
we obtain
\begin{equation}
W = \frac{\rho(1+w)}{3M^{6}}[2\rho(2+3w)+\sigma(1+3w)] \, ,
\end{equation}
which coincides with the result of \cite{CS}.

\section{Conclusions}\label{conclusions}

In this work, we considered general cosmological solutions and their stability
with respect to homogeneous and isotropic perturbations in the braneworld
theory with the induced-curvature term in the action for the brane. In our
approach, either the initial conditions or the constants of the theory were
perturbed. Part of the results are similar to those obtained by Campos and
Sopuerta \cite{CS} for the Randall--Sundrum model.  Specifically, the expanding
de~Sitter solution is an attractor, while the expanding Friedmann solution is a
repeller in the phase space of the theory. However, in the model with $m \ne
0$, there are no analogues of the solution considered in \cite{BDEL} (case~4 of
Sec.~\ref{method}), and an expanding universe dominated by dark radiation
(case~5 of Sec.~\ref{method}) is replaced by a somewhat different regime
(cases~2 of Secs.\@ \ref{generic} and \ref{special}), in which dark radiation
contributes similarly to matter with negative pressure $p = - \rho / 3\,$ and
to spatial curvature.

The possibility of static solutions, including those dominated by negative dark
radiation, was previously discussed in \cite{CS,GM} in the context of the
Randall--Sundrum model.  In the braneworld theory with the scalar-curvature
term in the action for the brane, static solutions with matter satisfying the
strong energy condition exist not only with closed spatial geometry but also
with open and flat ones even in the case where dark radiation is absent, as
well as in the case where it is negative. This brings to attention an
interesting possibility that the braneworld universe, even being spatially open
or flat, could have passed through a quasi-static (or ``loitering'') phase, the
details of which are further investigated in \cite{loiter}.

\section*{Acknowledgments}

D.~I.\@ is grateful to Dr.~Vitaly Shadura for creating wonderful atmosphere at
the Scientific and Educational Center of the Bogolyubov Institute for
Theoretical Physics in Kiev, Ukraine.

\end{document}